\documentclass[preprint2]{aastex}

\usepackage{subfig}
\newcommand{\msun}{\mbox{M$_{\odot}$}}

\shorttitle{Variability of massive LMC stars}
\shortauthors{Szczygie{\l} et al.}

\begin{document}

\title{Variability of Luminous Stars in the Large Magellanic Cloud
       Using 10 Years of ASAS Data}

\author{D. M. Szczygie{\l} and K. Z. Stanek}
\affil{Department of Astronomy, The Ohio State University, 140 West 18th Avenue,
Columbus, OH 43210, USA\\e-mail: szczygiel, kstanek@astronomy.ohio-state.edu}
\author{A. Z. Bonanos}
\affil{Institute of Astronomy \& Astrophysics, National   Observatory of Athens,
I. Metaxa \& Vas. Pavlou St.,\\ P. Penteli, 152 36 Athens, Greece\\e-mail: bonanos@astro.noa.gr}
\author{G. Pojma\'nski and B. Pilecki}
\affil{Warsaw University Astronomical Observatory,
Al. Ujazdowskie 4, 00-478 Warsaw, Poland\\e-mail: gp, pilecki@astrouw.edu.pl}
\and
\author{J. L. Prieto\altaffilmark{1}}
\affil{Carnegie Observatories, 813 Santa Barbara Street, Pasadena, CA 91101, USA\\
email: jose@obs.carnegiescience.edu}
\altaffiltext{1}{Hubble and Carnegie-Princeton Fellow}

\begin{abstract}
Motivated by the detection of a recent outburst of the massive luminous blue
variable \mbox{LMC-R71}, which reached an absolute magnitude $M_{V} = -9.3$ mag, we
undertook a systematic study of the optical variability of 1268 massive stars
in the Large Magellanic Cloud, using a recent catalog by \citet{Bonanos09} as
the input.
The ASAS All Star Catalog \citep{Pojmanski02} provided well-sampled light
curves of these bright stars spanning 10 years. Combining the two catalogs
resulted in 599 matches, on which we performed a variability search. We
identified 117 variable stars, 38 of which were not known before, despite their
brightness and large amplitude of variation.
We found 13 periodic stars that we classify as eclipsing binary (EB) stars,
eight of which are newly discovered bright, massive eclipsing binaries composed
of OB type stars.
The remaining 104 variables are either semi- or non-periodic, the majority (85)
being red supergiants. Most (26) of the newly discovered variables in this
category are also red supergiants with only three B and four O stars.
\vspace{1cm}
\end{abstract}


\section{INTRODUCTION}

Studies of stellar variability in the Magellanic Clouds have a long and
venerable history, starting with a 1904 Harvard College Observatory
Circular by E. C. Pickering stating that ``The two Magellanic Clouds
have long been objects of careful study, on account of the extraordinary
physical conditions which prevail in them. They have not, however,
heretofore been known as regions in which variable stars are numerous''
\citep{Pickering04}. That circular, although formally authored by
Pickering, in fact reported 152 new variable stars in the Large
Magellanic Cloud found by Henrietta Leavitt, work that was soon followed by
``843 New Variable Stars in the Small Magellanic Cloud''
\citep{Leavitt05}, ``1777 Variables in the Magellanic Clouds''
\citep{Leavitt08} and culminated in the formulation of the Cepheid
period--luminosity relation \citep[the ``Leavitt Law'',][]{Leavitt12},
a result of fundamental and continued importance in astronomy and
cosmology \citep[e.g.,][]{Freedman01, Macri06}.

Later studies of stellar variability in the Clouds included many
papers on Cepheids \citep[e.g.,][]{Caldwell86, Madore91}, but there were
also studies of long period variables \citep[e.g.,][]{Wood83, Feast89},
irregular variables of the S Doradus type
\citep[luminous blue variables, LBVs; e.g.,][]{Stahl83} and RR Lyr stars
\citep[e.g.,][]{Thackeray58, Walker92}, just to name the most popular
subjects. By virtue of being so close to us and being relatively unobscured,
the Magellanic Clouds continue to provide an ideal terrain for stellar
variability studies.

The idea of \citet{Paczynski86} to use the Magellanic Clouds as targets
for microlensing surveys to detect dark matter in the form of MACHOs has led
to a new and glorious era in variability studies in general, not just in
the Clouds, and has resulted in a wide variety and large number of
papers reporting on variability in the Large Magellanic Cloud (LMC) and
the Small Magellanic Cloud (SMC), including work
on Cepheids \citep[e.g.,][]{Alcock95, Udalski99, Soszynski08}, eclipsing
binaries \citep[e.g.,][]{Alcock97, Wyrzykowski04}, RR Lyr stars
\citep[e.g.,][]{Alcock96, Soszynski03, Soszynski09}, all kinds of stellar
variability \citep[e.g.,][]{Cook95, Zebrun01}, and even quasars
\citep[e.g.,][]{Geha03, Dobrzycki03, Kozlowski10}, supernovae
\citep[e.g.,][]{Garg07} behind the Clouds and supernovae light echoes
in the LMC \citep{Rest05}. Some microlensing events were also found in the
Clouds, but clearly not enough to populate the halo of our Galaxy
with MACHOs \citep[e.g.,][]{Alcock00, Tisserand07, Wyrzykowski09}.

Given all that tremendous effort and many hundreds of papers written
about stellar variability in the Magellanic Clouds, is there still room
for more discoveries? As is almost always the case in astronomy, the answer
is ``yes''. One path is simply to go deeper, and some impressive results
have been reported recently, such as the detection of extragalactic W
UMa binaries in the LMC \citep{Kaluzny06} using ground-based data from
a large aperture telescope. Another path is to find variable stars in the
densest parts of the Clouds, which are now accessible using the {\em Hubble
Space Telescope} \citep[e.g.,][]{Fiorentino08}. Still yet another path is
to take advantage of the fact that we have been observing stars in the
Magellanic Clouds for a long time now \citep[e.g.,][]{Kozlowski10}, even
as long as a hundred years \citep{Grindlay09}.
But the path taken in this paper, namely to go {\em shallower}, and to cover
all of the Clouds (the SMC will be discussed in a future paper), again owes
its existence to Bohdan Paczy\'nski, whose vision also led to the All Sky
Automated Survey (ASAS) project led by Grzegorz Pojma\'nski, who created
it and has kept it functional for the past 10 years.

\begin{figure*}
\begin{center}
\begin{tabular}{c}
\includegraphics[width=0.99\linewidth]{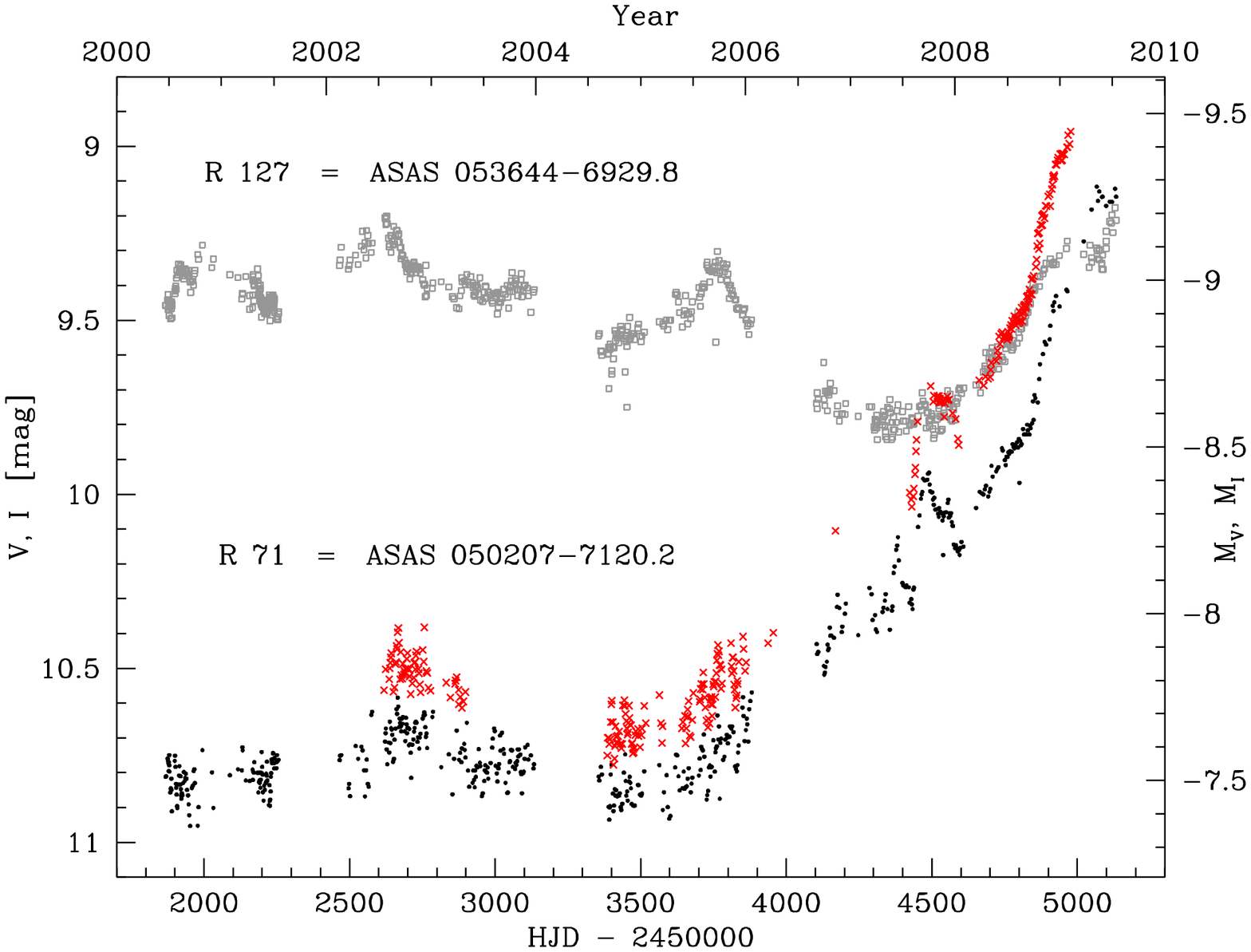}
\end{tabular}
\end{center}
\vspace{-0.5cm}
\caption{ASAS light curves of the LBVs R 71 in $V$ (black circles) and $I$
(red crosses) bands, and R 127 in $V$ (gray open circles), spanning 10 years.
The right-hand axis shows the absolute magnitude value, assuming an LMC
distance modulus of 18.41 mag \citep{Macri06}.}
\label{r71}
\end{figure*}

A particular motivation for our study was provided by the recently
reported outburst of an LMC LBV R71\footnote{R71 = ASAS 050207--7120.2},
which during the last 5 years has brightened by more than 2 magnitudes to a
visual magnitude of about 9.0, becoming the brightest star in the LMC
\citep{Gamen09}.  As it turns out, the ASAS 
\citep{Pojmanski97, Pojmanski02, Pojmanski05} photometric
database contained a detailed $V$-band light curve of that star (see
Figure \ref{r71}), obtained and reported in an automatic fashion
by ASAS. Motivated by the fact that the brightening of R71 started several
years ago, but was first reported only in 2008, we decided to investigate
light curves of other bright stars in the LMC, using the ASAS data.
R127 is another example of a ninth magnitude LBV in the LMC that underwent
dramatic change starting in 2005 that was recorded by
ASAS\footnote{R127 = ASAS 053644--6929.8 (although blended)} (see Fig.
\ref{r71}), and was not noticed until two years later by \citet{Walborn08}, who
began monitoring it spectroscopically. There are bound to be other such examples
of bright LMC variables already recorded in the ASAS database. We therefore used
the recent catalog of \citet{Bonanos09} as an input catalog to characterize
the variability properties of these known LMC massive stars. Confirming our
suspicion, there are not only many spectacular light curves among known
variables, but also a significant fraction of new interesting objects.

This paper has the following structure. In Section 2, we briefly describe
both the input catalogs and the matching procedure. Section 3 introduces our
variable star selection methods and discusses the identified variables.
In Section 4, we present color--magnitude diagrams (CMDs) of the variable stars
and discuss special cases. We summarize this work in \mbox{Section 5.}

\section{DATA}

\subsection{Input Catalogs}

We use the catalog of \citet{Bonanos09}, hereafter B09, as the source of
bright, massive objects associated with the LMC. It
contains photometric measurements for 1268 massive stars in several visual and
infrared wavelengths ({\em UBVI, JHK$_{S}$}, and 3.6-24 $\mu$m), as well as their
spectral classifications. The area covered by this catalog is delimited by
$-72^{\circ} < \delta < -64^{\circ}$ and $70^{\circ} < \alpha < 90^{\circ}$.

The ASAS\footnote{http://www.astrouw.edu.pl/asas}
\citep{Pojmanski02} has been monitoring the southern sky ($\delta <
+28^{\circ}$) since 2000 and by now the average number of data points per star
in the LMC region is 550 in the $V$ band and 350 in the $I$ band. The $V$
magnitude range covered by the ASAS telescopes is around 8 - 14.5, which ideally
suits a study of the brightest objects in the LMC.  The number of objects
visible in the LMC direction, within the coordinates listed above, is about
28 000, many of them being Galactic field stars.

\subsection{Matching Procedure}

The main difficulty in combining the two catalogs resulted from the large size
of the ASAS pixel, which is $\sim$14\farcs9. Thus, even though coordinates of
objects in the ASAS catalogs are accurate to $\sim 3^{\prime\prime}$
\citep{Pojmanski97} and the positions within the B09 catalog are typically
accurate to less than $\sim 1^{\prime\prime}$, in very crowded regions one ASAS
object may be composed of more than one (bright) LMC star.

For this reason we performed catalog matching in two steps. First, each of the
1268 stars from the B09 catalog was assigned one closest counterpart from the
ASAS catalog (28,000 sources), without putting any constraints on counterpart
separation. Among 1268 matching entries, 761 turned out to be unique,
which means that an ASAS star in such a pair was assigned only one counterpart
from the B09 catalog. Within the remaining 507 matching entries, one ASAS
star was assigned to either two or three B09 stars and as a result these 507
matching entries contained only 162 unique ASAS sources. This was the
case in the densest regions of the LMC, where due to low ASAS resolution
objects from the B09 catalog could not be resolved.

In the second step we took the 923 (761 + 162) ASAS stars selected in the
previous step and checked how many B09 neighbors each ASAS star had within
a radius of $15^{\prime\prime}$ (1), between radii $15^{\prime\prime}$ and
$30^{\prime\prime}$ (2), and between $30^{\prime\prime}$
and $45^{\prime\prime}$ (3).
Given that the aperture radius adopted by us is in most cases ($\sim$70\%)
1 ASAS pixel, we assumed that the match could be considered unambiguous when
there was only one neighbor within (1) and none within (2). There were 675
objects fulfilling this criterion. However, for about 30\% of ASAS stars we
have adopted apertures greater than 1 pixel, which could result in neighboring
stars  overlapping each other, so we decided to apply a more conservative
criterion, that no stars were allowed in either zone (2) or (3). The number
of matches was reduced to 599, which we accepted as our final sample.
The choice of the aperture size described above is automatic and depends only
on the magnitude of an object --- for the faintest stars (13 - 14 mag) we
use the smallest aperture radius of 1 pixel, while for the brightest
(8 mag) we use the largest aperture radius of 3 pixels.

It is important to stress here that the match has been made only against the
B09 catalog, which means that the foreground field stars are not accounted for.
At the same time, the B09 catalog itself is far from a complete list of massive
stars in the LMC, as it contains only the stars for which spectra exist.
Therefore, we should expect some crowding problems even in our final sample.
This issue is addressed in Section 3.3.

\begin{figure*}
\begin{center}
\includegraphics[width=0.99\linewidth]{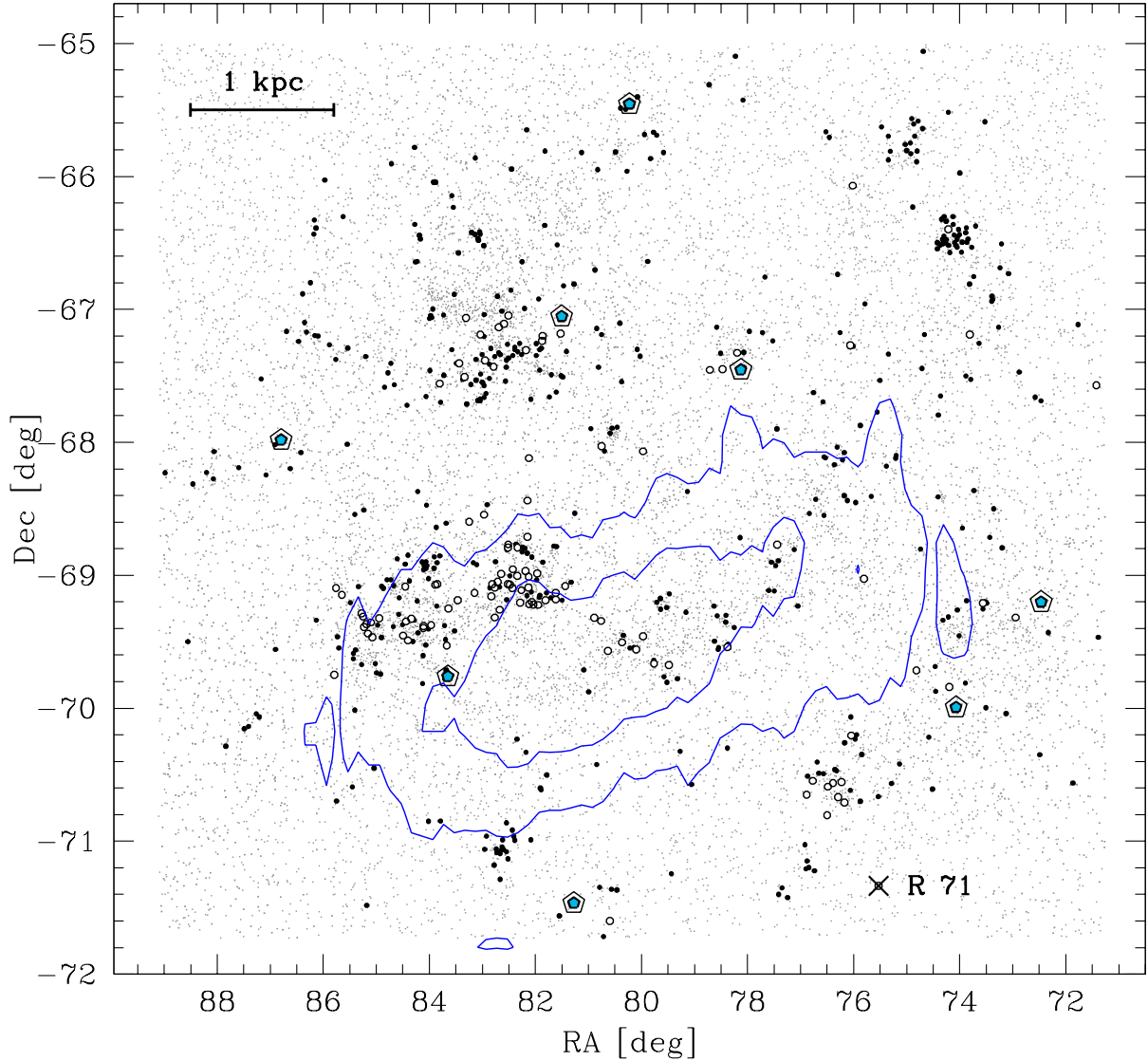}
\end{center}
\vspace{-0.5cm}
\caption{Spatial distribution of ASAS stars in the LMC region in equatorial
coordinates. Gray dots are the $\sim$19,000 objects brighter than $V=14$ mag
(out of $\sim$28,000 total) found by ASAS in this region of the sky.
Filled circles represent the 599 final matches between the ASAS and B09 catalogs
and open circles are the subset that was identified as variable (117 stars).
Blue pentagons mark new EB candidates (see Section 3 for details).
To show the location of the LMC bar, we also plot the stellar density contours
based on the OGLE-III photometric maps of the LMC \citep{Udalski08}.
We only include $I<15$ mag objects, binned into 0.2 deg regions in both
R.A. and Decl. The contour levels are 50 (for the outer) and 90 (for the inner)
stars per bin. The outer contour includes the 30 Doradus star-forming region.
Note that the OGLE fields do not cover the Shapley Constellation III
star-forming region.
The 1 kpc marker is based on 18.41 mag LMC distance modulus \citep{Macri06}.}
\label{radec}
\end{figure*}

The spatial distribution of the 599 matched objects in equatorial coordinates
is presented in Figure \ref{radec} with black filled and open circles.
Gray dots represent all ASAS objects in this region brighter than $V=14$ mag
($\sim$19,000 of $\sim$28,000 points). The sample is dominated by foreground
stars.

\section{RESULTS OF VARIABLE STAR SEARCH}

As already summarized in the Introduction, the LMC has been the target of
variability searches since the beginning of the last century, with an
increasing intensity over the last two decades. Recent catalogs of variable
stars were mainly generated using the data acquired by microlensing surveys such
as OGLE and MACHO.  However, these catalogs are complete for stars fainter than
13 - 14 $V$ magnitude, and are missing the brightest objects. This oversight
can be corrected with the aid of ASAS data.

\subsection{Variable Star Selection}

In order to investigate variability of the 599 matched stars, we extracted
$V$-band light curves from the online ASAS database, keeping only points with
quality flag ``A''. Then we cleaned the light curves by removing points that
lie more than 3$\sigma$ from a local linear model fitted to each light curve,
where $\sigma$ is the standard deviation.
All cleaned light curves were subject to variability analysis using the
analysis of variances (AoV) algorithm \citep{Schwarzenberg89} in the frequency
range $0.0001 - 8$ c d$^{-1}$. Each light curve was then folded with a period
equivalent to the frequency with the highest peak in the power spectrum and
visually inspected to eliminate possible spurious detections such as 1 day
aliases. During the visual inspection other frequencies with high signal were
examined and the best period was chosen.

\begin{figure*}
\begin{center}
\includegraphics[width=0.85\linewidth]{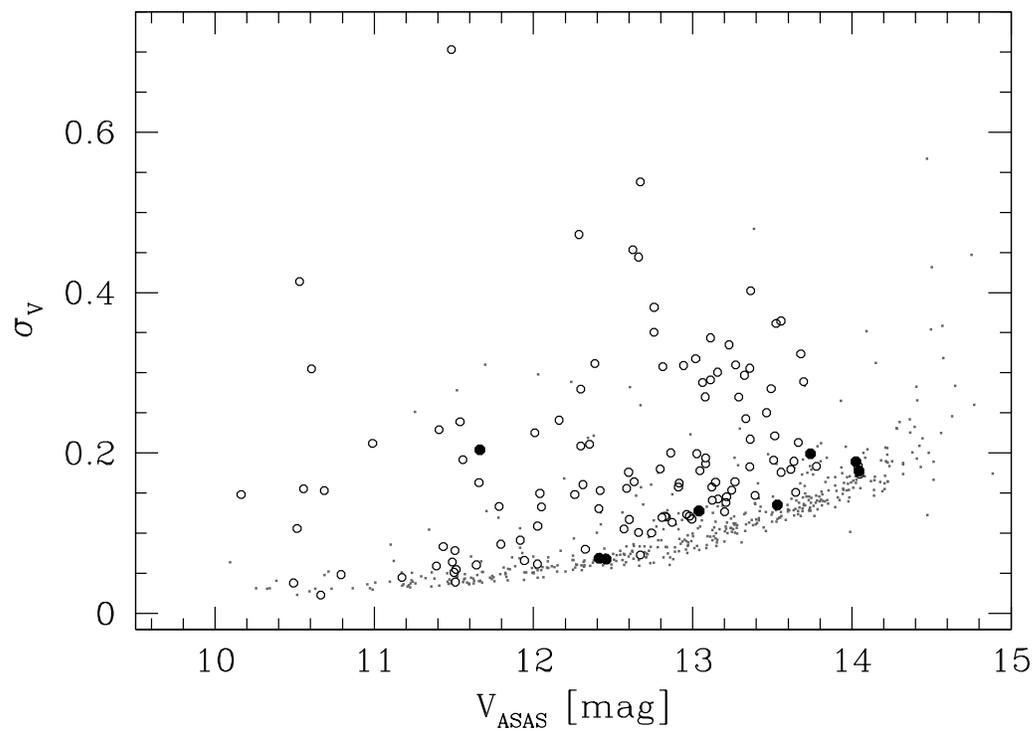}
\end{center}
\vspace{-0.5cm}
\caption{Standard deviation $\sigma_{\rm V}$ vs. $V$-band mean magnitude for
all 599 matched sources. Variable stars are marked with open circles; EB
candidates (eight points) are denoted by filled circles (see Section 3.1
for details).}
\label{rms}
\end{figure*}

Figure \ref{rms} presents the standard deviation $\sigma_{\rm V}$ versus $V$-band
mean magnitude with all 599 matched sources marked with gray dots and variable
stars with larger circles. There are a number of stars with high
$\sigma_{\rm V}$ values that were not classified as variable. These objects
have an increase in brightness around HJD 2452200 and 2452600 which was due to
 strong defocusing of the ASAS telescopes.

During the inspection process we identified 117 variable stars, of which 55 
were not listed as variable in the Simbad\footnote{http://simbad.u-strasbg.fr}
database. We then matched these 55 stars against the GCVS catalog
\citep{Samus09} using VizieR\footnote{http://vizier.u-strasbg.fr} and found
that another 24 were known to be variable, which altogether gives 79 known
variable stars and 38 new discoveries.

\subsection{Periodic Variables}

Among our 117 detections there are 13 periodic variables; five of them are
known binary systems (although BAT99--32 was only known as a spectroscopic
binary and its eclipsing nature has not been reported before), which are as
follows:

\begin{itemize}

\item 045515--6711.4 = BAT99--6 = Brey 5 = Sk --67 18 \citep{Prevot81}:
  \citet{Seggewiss91} have identified this object as a binary system
  with a tentative period of 1.99 days, based on a very poor light curve. Later
  \citet{Niemela01} have shown it to be ``a multiple system consisting of at
  least two pairs of short period binaries.'' We confirm the 2 day period of
  the eclipsing pair.

\item 052223--7136.0 = BAT99--32: a long known Wolf--Rayet (W-R) star
  \citep{Westerlund64}, which has been reported to be a single lined
  spectroscopic binary \citep{Moffat89} and an EB with a period
  of $\sim$1.91 days \citep{Seggewiss91}. The discovery from ASAS
  observations that it is also an eclipsing system with a period
  $P\sim$3.8 days that is twice as long as that found in the
  literature, explains the low radial velocity amplitude and unusually
  short period previously reported. High resolution spectra of this
  system would be very valuable as they would allow for a determination
  of its parameters, given that very few W-R stars have good
  measurements of masses and radii.

\item 052606--6710.9 = Sk --67 105: \citet{Niemela86} found it to be a
  massive double lined spectroscopic binary (the first extragalactic
  binary to be studied), while \citet{Haefner94} discovered it is a near
  contact eclipsing system. \citet{Ostrov03} derived component masses of
  $48.3\,\msun$ and $31.4\,\msun$.

\item 052727--6712.0 = Sk --67 117 = HV 2543: a known semi-detached
  double-lined EB with masses $25.63\pm0.7\,\msun$ and
  $15.63\pm1.0\,\msun$ \citep{Niemela94, Ostrov00}.

\item 053442--6931.6 = MACHO81.8763.8: discovered as an EB
  by \citet{Alcock97}, it was shown to be a massive binary with
  component masses of $41\,\msun$ and $27\,\msun$ by \citet{Ostrov01}.

\end{itemize}

\begin{figure*}
\begin{center}
\begin{tabular}{c}
\includegraphics[width=0.65\linewidth]{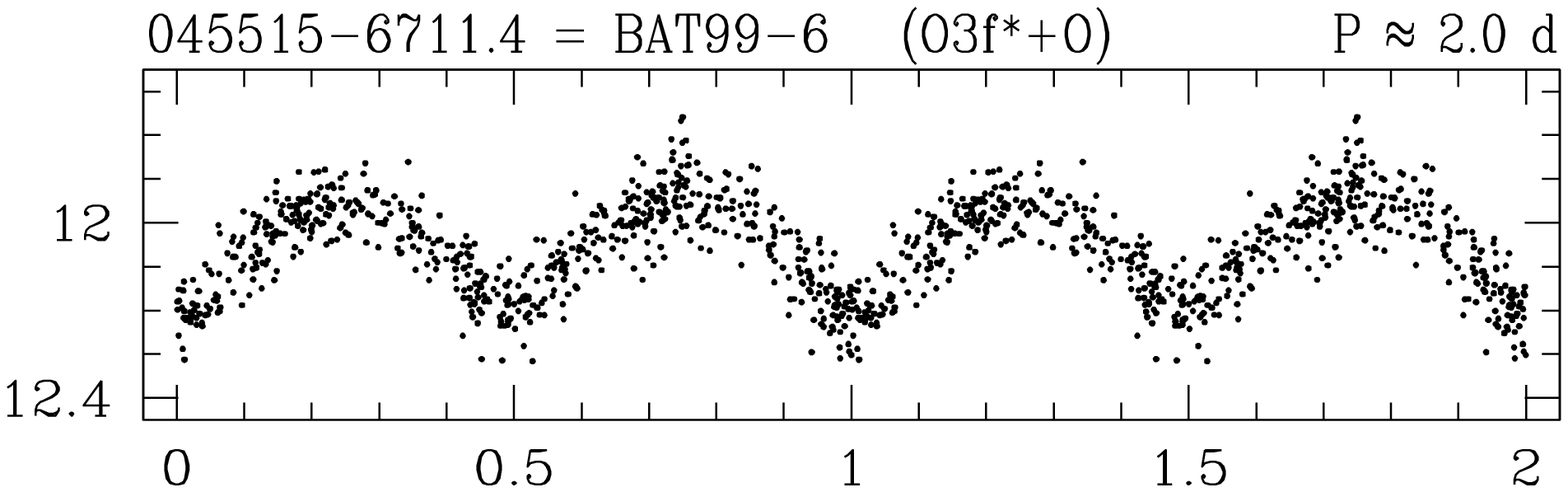} \\ [-2.8ex ]
\includegraphics[width=0.65\linewidth]{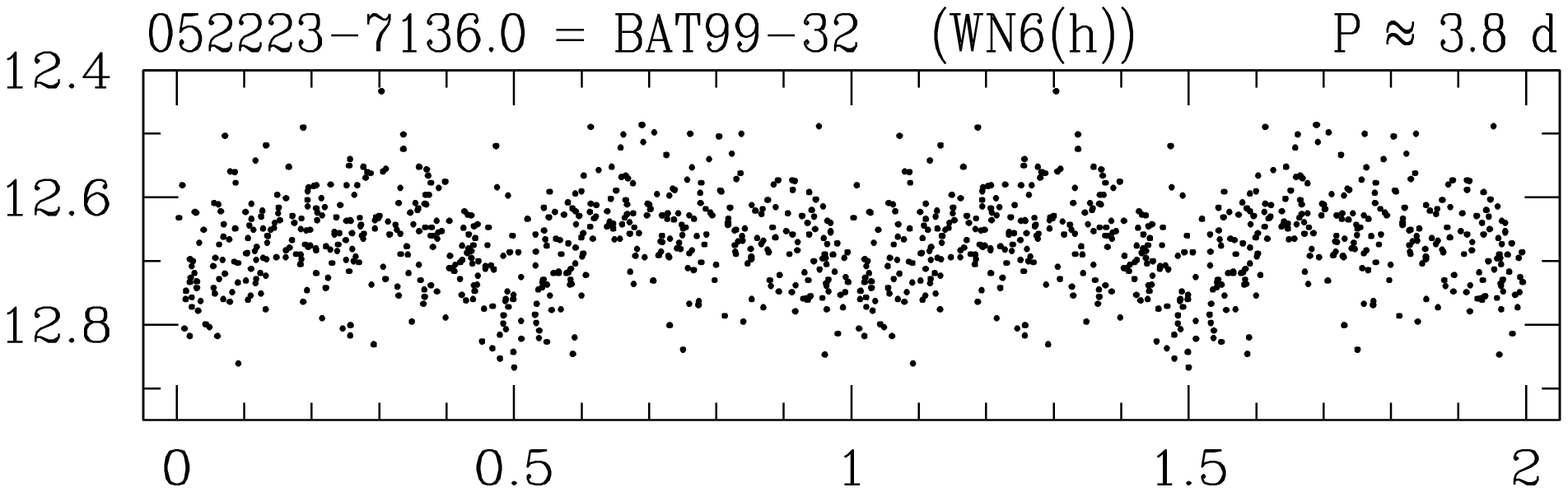} \\ [-2.8ex ]
\includegraphics[width=0.65\linewidth]{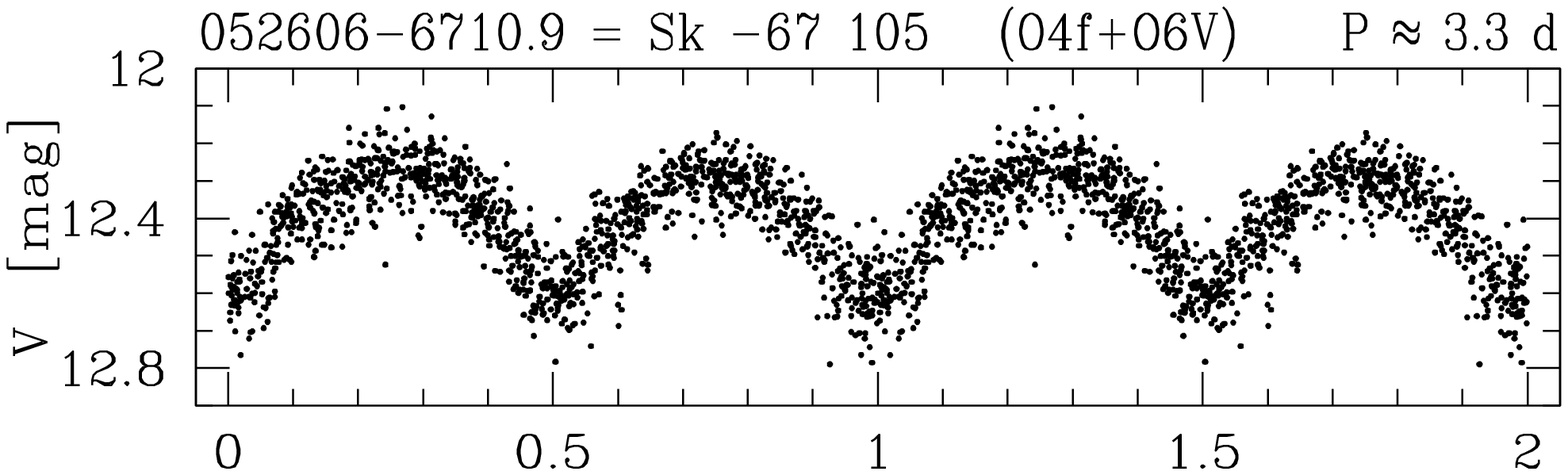} \\ [-2.8ex ]
\includegraphics[width=0.65\linewidth]{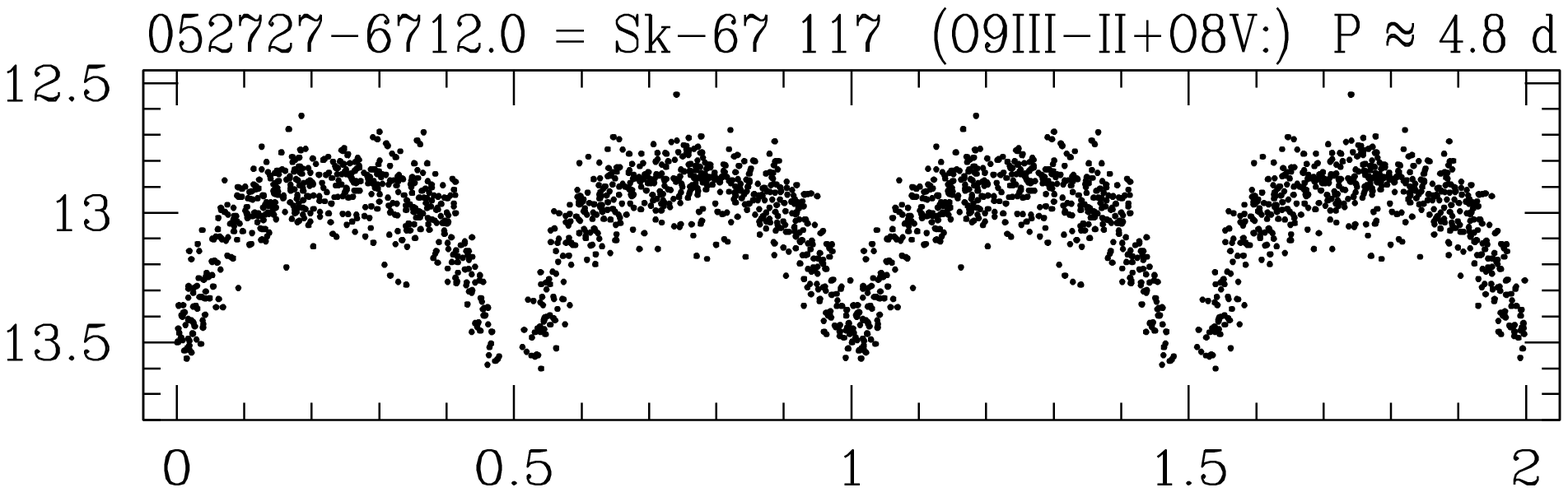} \\ [-2.8ex ]
\includegraphics[width=0.65\linewidth]{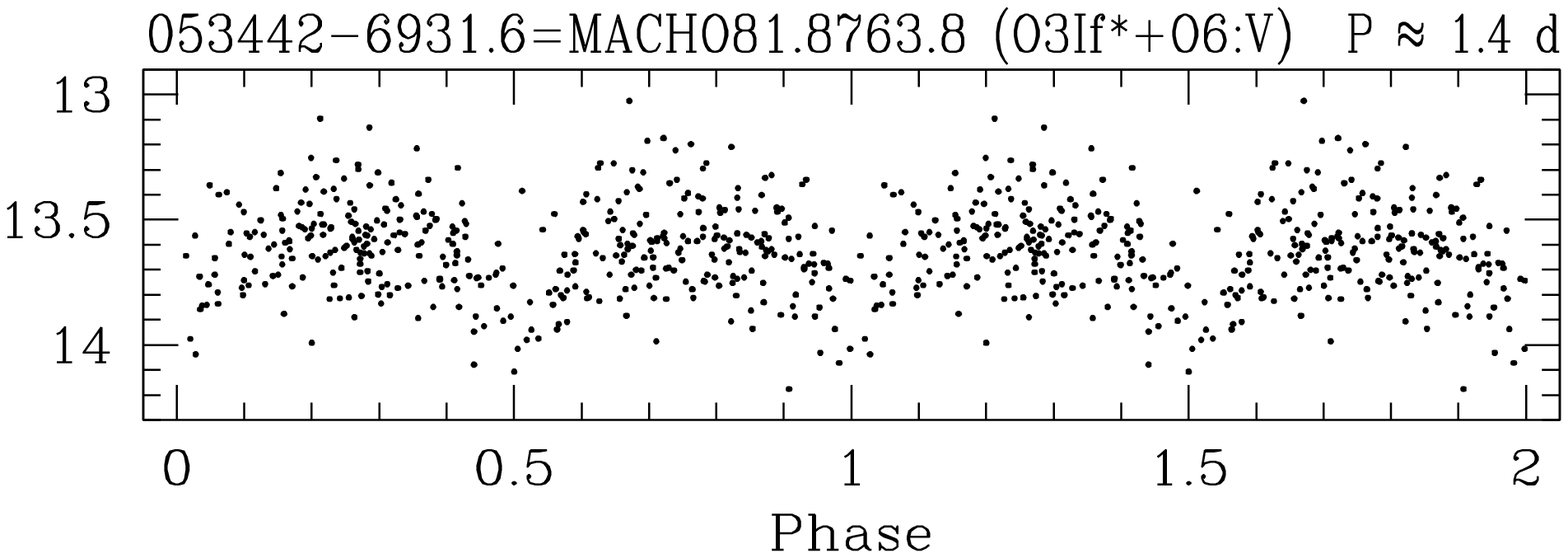} \\
\end{tabular}
\end{center}
\caption{Phased ASAS $V$-band light curves of the five previously known
EBs, sorted by right ascension, spanning 10 years. The header of each light
curve contains two IDs, spectral type (in brackets), and orbital  period
in days. For more information on these stars, see Table 1.}
\label{5eb}
\end{figure*}

The $V$-band light curves of these binaries, spanning 10 years of ASAS
observations are plotted in Figure \ref{5eb} and their parameters are presented
in Table 1. In combination with previously published photometric data,
the ASAS observations can be used to improve the precision of the binary
orbital periods, by extending the baseline of the observations.

\begin{figure*}
\begin{center}
\includegraphics[width=0.99\linewidth]{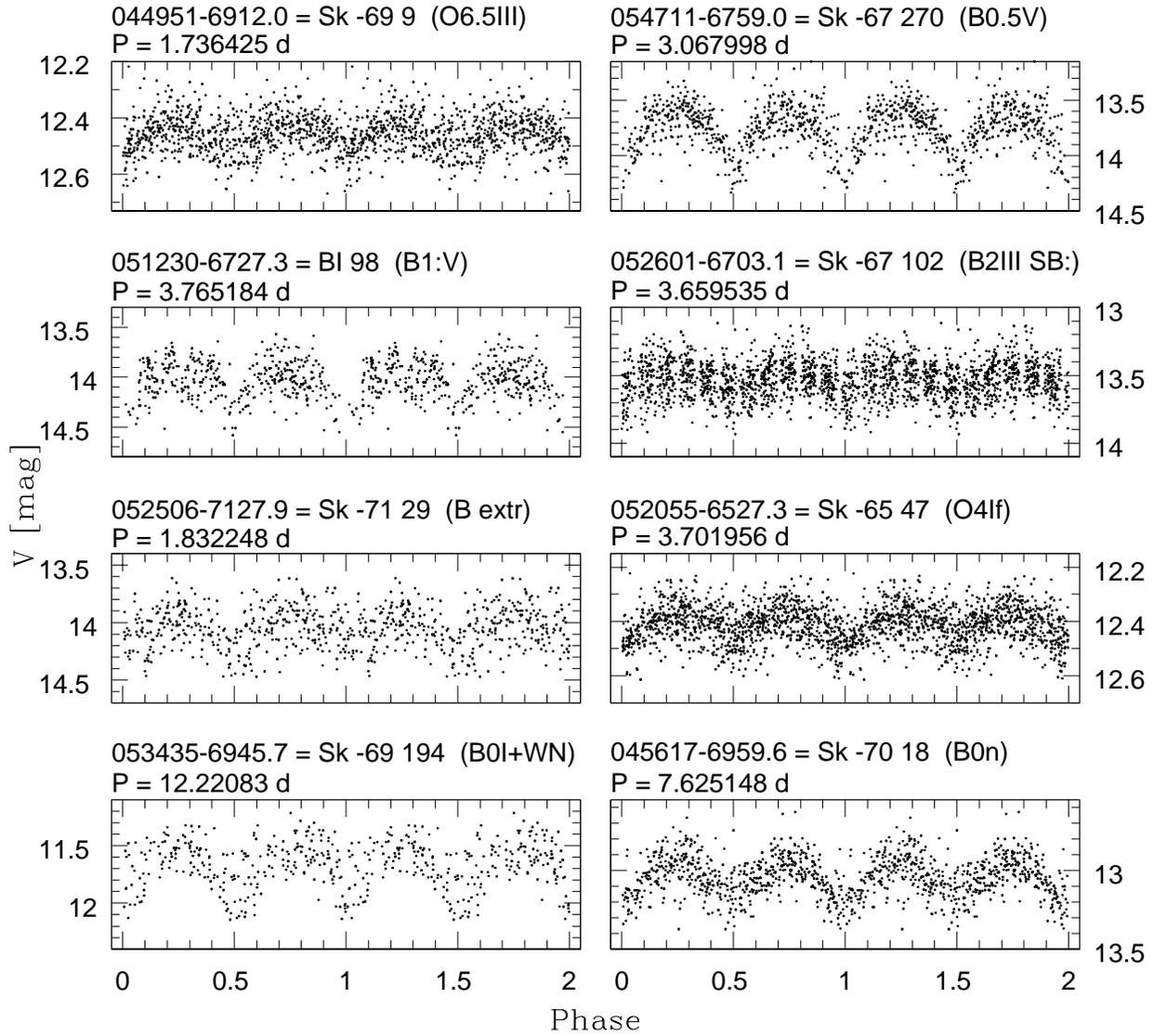}
\end{center}
\caption{Phased ASAS $V$-band light curves of eight new EB candidates,
sorted by right ascension. The header of each light curve contains two IDs,
spectral type (in brackets), and orbital period in days. For more information
on these stars, see Table 2.}
\label{8eb}
\end{figure*}

Parameters for the remaining 8 objects (blue pentagons in Figure \ref{radec})
are listed in Table 2, and their light curves are plotted in Figure \ref{8eb}.
They are most probably newly discovered EB stars and we have
already started follow up spectroscopic observations in order to verify their
binarity. All have OB spectral types, which is consistent with their short
orbital period. Three of them had been recognized as variable stars long ago,
one was misclassified as a long period variable (052506--7127.9 = Sk --71 29 =
LMC V2528\footnote{``Poorly studied irregular variables of early (O-A)
spectral type.'' GCVS}) by \citet{Hughes89}, and the other two had no type
specification: 051230--6727.3 = BI 98 \citep{Butler74} and 053435--6945.7 =
Sk --69 194 \citep{Westerlund61}.

The latter, namely 053435--6945.7 (Sk --69 194) has the longest period of all
EB candidates ($P=12.2$ days). According to \citet{Massey00} it is a B0 Ia + WN
star.  The filling-in of the minima in its light curve (Figure \ref{8eb}) most
probably results from the presence of a bright ($V\approx11$ mag) foreground
K type star HD 269770, distant by $33^{\prime\prime}$ in the SW direction.

The remaining five EB candidates were not previously identified as variable
stars. Their ASAS identifications are as follows:

\begin{itemize}

\item 045617--6959.6 = Sk --70 18 \citep[B0 n;][]{Conti86}

\item 044951--6912.0 = Sk --69 9 \citep[O6.5 III;][]{Jaxon01}

\item 052055--6527.3 = Sk --65 47 (O4 If), whose spectrum is available in
  the literature \citep[Figure 7 of][]{Massey05}. The He II $\lambda 4686$ and
  H$\alpha$ lines are not well fit by the model, which may be due to wind
  interaction in the binary. The object was recently reclassified to
  O4 I(n)f+p type by \citet{Walborn10}.  Since O4 If stars are quite rare,
  measuring the parameters for this system would be valuable.

\item 052601--6703.1 = Sk --67 102 \citep[B2 III;][]{Conti86}

\item 054711--6759.0 = Sk --67 270 \citep[B0.5 V;][]{Conti86}

\end{itemize}

\subsection{Non-periodic Variables}

We identified 104 variable stars that are either semi-periodic or
non-periodic. Among them are 25 variables that were already listed in
the ASAS Catalog of Variable Stars (ACVS) as MISC-type variable stars, but
9 of these are not listed as variable in the Simbad database. Altogether
there are 33 objects that are not marked as variable in Simbad, GCVS,
or ASAS.
All of them are bright and many have large light curve amplitudes, so it is
noteworthy that their variability had not been discovered previously. The
reason that some of them do not figure in ACVS is that the part of the catalog
containing the LMC region was made when the light curves consisted of 40--50
observations only \citep{Pojmanski02}, during which there was no significant
change in the light curves.

\begin{figure*} 
\begin{center}
\includegraphics[angle=-90,width=0.99\linewidth]{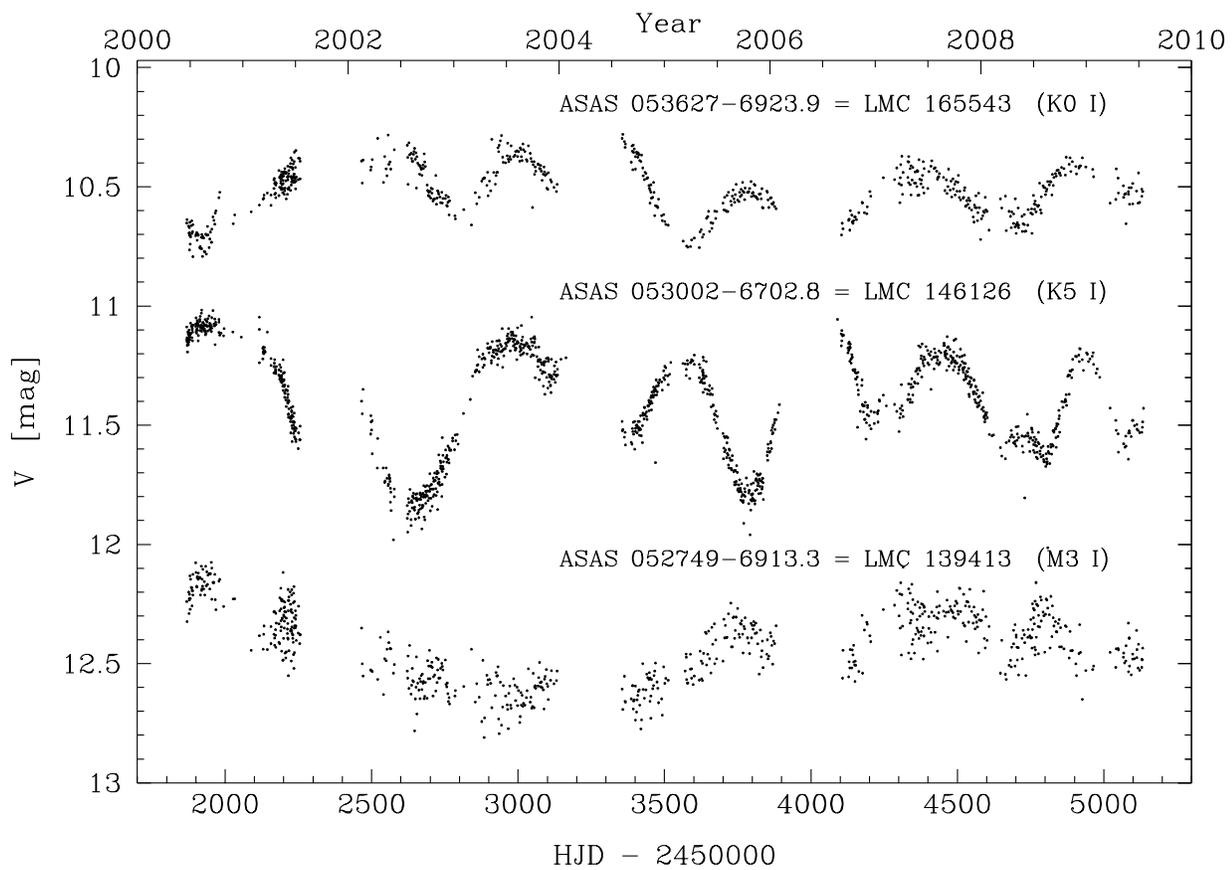}
\end{center}
\caption{Light curves of 3 RSG out of 38 objects that are newly reported as
variable (ASAS 052749--6913.3 = 2MASS J05274747--6913205, ASAS 053002--6702.8
= 2MASS J05300226--6702452, ASAS 053627--6923.9 = 2MASS J05362678--6923514 =
CPD-69 421), spanning 10 years. The spectral type from B09 is given in brackets.
}
\label{3rsg}
\end{figure*}

\begin{figure*}
\begin{center}
\includegraphics[width=0.99\linewidth]{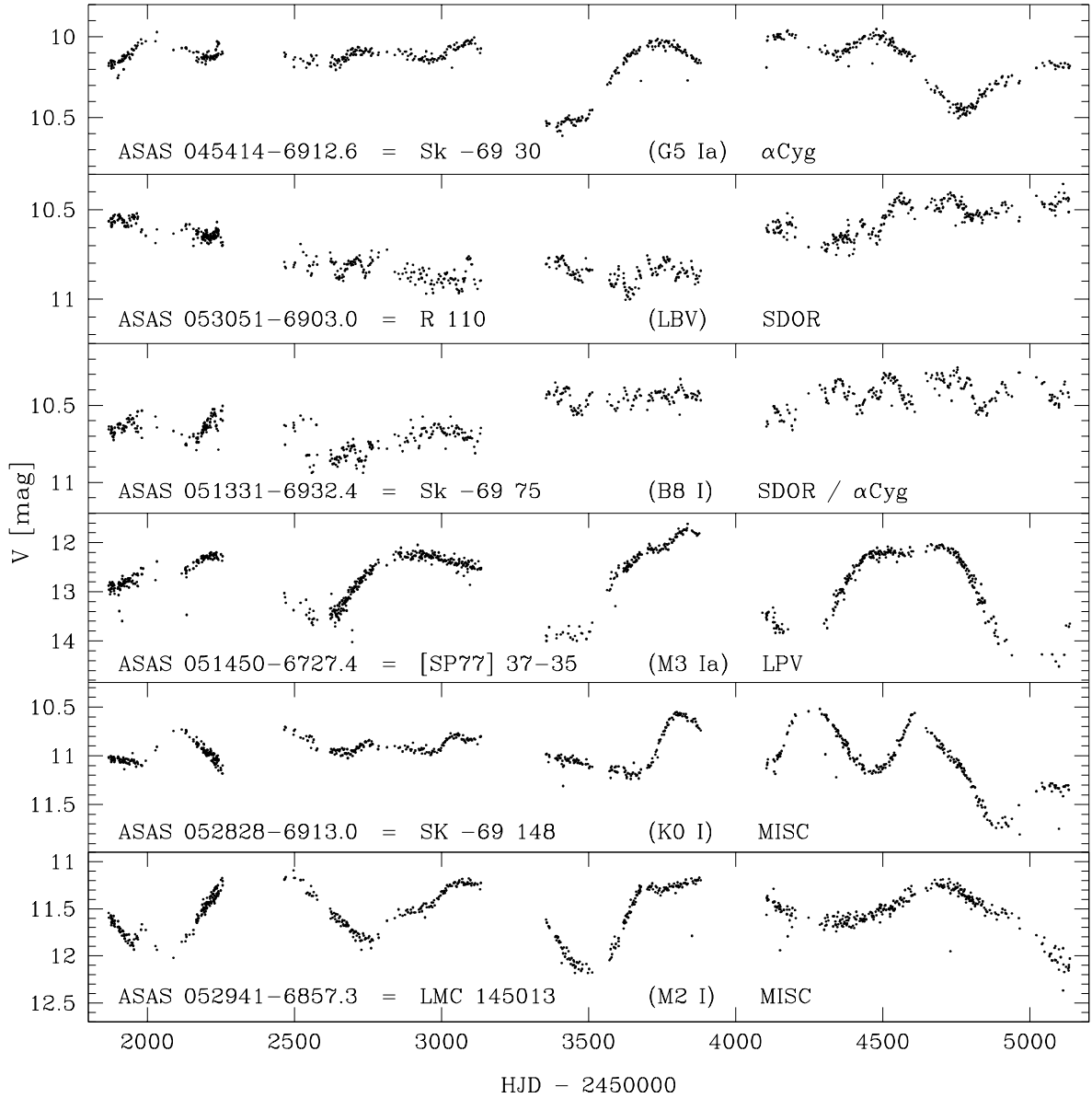}
\end{center}
\caption{Ten-year long light curves of a few interesting known variable stars.
The spectral type is given in brackets, followed by the variability type as found
in the literature. The variability of the last two objects was discovered by
ASAS \citep{Pojmanski02} and classified as ``Miscellaneous'' (MISC).}
\label{ex_lc}
\end{figure*}

The light curves of some interesting objects that are for the first time
reported as variable are plotted in Figure \ref{3rsg}, while Figure \ref{ex_lc}
shows long term light curves of a few known interesting variables; the last two
are solely ASAS discoveries. In Figure \ref{104_lc} we present light curve
stamps of all 104 semi- or non-periodic stars.

The parameters of all 117 variable stars, both periodic and non-periodic are
listed in Table 3, supplemented with extinction values from \citet{Pejcha09}
where available (101 objects). The next to last column contains information on
whether the star is a known variable (by providing its ID) or if it is a new
discovery (empty field).\\
The last column provides information about crowding problems mentioned in
the fourth paragraph of Section 2.2. It contains the number of bright
companions and their approximate angular distance (in arcsec) from the variable
star. These objects were selected by examining the Digitized Sky Survey (DSS)
images around each variable star, using the Goddard 
{\it SkyView}\footnote{http://skyview.gsfc.nasa.gov/cgi-bin/query.pl} utility.
The DSS exposures reach 2--3 mag deeper than ASAS observations, making
them ideal for resolving objects otherwise indistinguishable by ASAS.

The search showed that 47 out of the 117 variables have fairly bright
close companions. Usually the presence of such neighbors will be reflected in
the light curves by a significant scatter in the observations, as in the case
of an EB Sk --69 194 already mentioned in Section 3.2. Another two examples are
RSG stars 053627--6923.9 (one bright star $\sim10^{\prime\prime}$ away) and
052749--6913.3 (1 very bright star $\sim20^{\prime\prime}$ away) presented
in Figure \ref{ex_lc}.

Table 3 is available as a whole in the electronic version of this article.

Using the classification of massive stars adopted in B09 we have divided the
stars in our sample into the following groups, sorted by the number of objects
in the group (the abbreviation is given in brackets).
For each class we list the number of variables, including periodic variables,
and the total number of stars in our matched sample:

- early B type stars (earlyB): 10 / 238,   

- O type stars (O): 10 / 156,              

- red supergiants (RSGs): 85 / 97,          

- late B type stars (lateB): 2 / 51,       

- Wolf Rayet stars (WR): 3 / 35,           

- A, F, G supergiants (AFG): 1 / 10,       

- supergiant B[e] stars (sgBe): 3 / 8,     

- luminous blue variables (LBVs): 3 / 3,    

- Be/X-ray binaries (bexrb): 0 /1.\\       
As expected, RSGs constitute the majority of all variable stars
(85/117=73\%), and they have the largest variability percentage among all
classes other than the LBVs (by definition). It is also difficult to confirm
that the 12 RSGs that we do not report as variable are truly not variable.
Their light curves show quite high scatter, which could indicate lower
amplitude variability than is detectable by ASAS. The percentage of variables
in other classes varies from about 4\% for B type stars to about 37\% for sgBe
stars, but the number of objects in several groups is too small to perform any
global statistics.

We will discuss some of these variable stars in the following section.

\section{VARIABLE STAR PROPERTIES}

\subsection{Color--Magnitude Diagrams}

In the left panel of Figure \ref{vi} we plot a CMD, showing $V$ versus
\mbox{$V-I$} magnitudes of all final matches between ASAS and B09 catalogs.
$V$ and $I$ are the mean magnitude values based on the ASAS light curves. Gray
open squares represent all 599 matches, while black dots stand for variable
stars. EBs are marked with big open circles.
The majority of variables have red colors, as noted in the previous section.
At the same time almost all red stars turn out to be variable, which supports
previous studies of RSG variability
\citep[e.g.,][and references therein]{Kiss06}.
All eclipsing variables are among the blue stars.
The right panel of Figure \ref{vi} presents $V$ amplitude vs \mbox{$V-I$} color
for the same sample with the same designations. The amplitude range is large,
from $\sim$0.1 to $\sim$1.8 mag.
Non-variable stars (gray squares) are plotted to show the scale of
observational scatter in the dense regions of the LMC. For these, the amplitude
$A_{\rm V}$ corresponds to the scatter in the data points, being roughly the
3$\times\sigma_{\rm V}$ value. The majority of objects with high $A_{\rm V}$
values are among the faint stars (see Figure \ref{rms}).

The star with the largest amplitude value, namely ASAS 051450--6727.4 =
[SP77]37-35 = HV 916, is a known variable supergiant of type M3 Ia (its light
curve is plotted in Figure \ref{ex_lc}), while R71 with its dramatic rise is
only fifth in amplitude and is marked with a cross in the right panel of
Figure \ref{vi}. The lowest amplitude variable is ASAS 045647--6950.4 = R66
which has been recognized as S Dor / $\alpha$ Cyg type variable by
\cite{vanGenderen02} and will be discussed in Section 4.2.

Figure \ref{36_45} is a CMD diagram of {\em Spitzer}/IRAC [3.6] versus
\mbox{[3.6]--[4.5]} infrared magnitudes of the 117 variables, taken from B09
catalog.
This figure is analogous to the Figure 2 of B09 where different spectral types
are marked with different symbols, but here we additionally mark eclipsing
variables with large open circles. Again we see that the majority of variable
stars are among RSGs, occupying the region of [3.6] $<$ 9 and
\mbox{[3.6]--[4.5]} $<$ 0.4 mag. All EBs are either O- or early B-type objects,
 except for one W-R star.

\subsection{Special Cases}

Three of the sgB[e] stars, marked with blue diamonds in Figure \ref{36_45},
were previously reported as variable by \citet{vanGenderen99, vanGenderen02}
and exhibit low amplitude variability:
S22 = HD 34664 = ASAS 051353--6726.9 ($\sim0.21$ mag),
R66 = ASAS 045647--6950.4 ($\sim0.08$ mag) and
R126 = 053626--6922.9 ($\sim0.16$ mag). We find evidence for a long-term
S Dor-like oscillation, on the order of six years, in S22, similar to that
reported by \citet{vanGenderen99}, who conclude that this star ``is not only
a B[e] star, but also a very weak-active LBV''.

In the case of R66 \citet{vanGenderen02} observed ``a long-term wave with
a possible timescale of hundreds of days and an amplitude of $\sim$0$\fm$06''
based on their observations from 1989 to 1991. This is the lowest amplitude
variable in our sample, with $A_{\rm V}=0.08$ being consistent with previous
estimates. We observe a pulsation-type oscillation with a period of 224 days,
which is roughly four times longer than the 55 day period observed by
\citet{vanGenderen02}, attributed to $\alpha$ Cyg-type oscillations. We do not
observe longer S Dor-like oscillations in this star.

\section{SUMMARY}

The aim of this paper was to investigate photometric variability of massive
stars in the LMC. We performed a variability search among
599 out of 1268 objects listed in the catalog of \citet{Bonanos09} which had
their unique counterparts in the ASAS \citep{Pojmanski02} database, spanning
10 years of observations.
We found 13 periodic and 104 semi- or non-periodic variables brighter than
$V=14$ mag, of which $\sim$32\% had not been listed as variable in either
Simbad, GCVS, or ASAS catalogs.

The majority of variable stars (73\%) identified are RSGs and almost
all RSGs exhibit some form of variability. The variability rate
among other spectral types fluctuates between $\sim$4\% (for O- and B-type
stars) and $\sim$37\% (for sgB[e] stars), but the absolute numbers of stars
in the groups are too small to perform reliable statistics.

All 13 periodic variables are EB stars, but to our surprise
only five of them were known to be binary before, even though the LMC had
been surveyed for bright variables about a century ago at Harvard. However, the
sensitivity of the photographic plates only allowed for detection of variables
with amplitudes above 0.5 mag, while all new EBs have amplitudes below 0.6 mag.
The eight new EBs have O and B spectral types, which are consistent
with their short orbital periods. We have already started follow-up
spectroscopic observations in order to verify their binary nature and
investigate their physical properties.

We present some of the more interesting 10 year long light curves, which for
many objects are the most extensive light curves available in the literature.
This is particularly important since several of the systems are relatively
rare. We also provide light curve stamps for all semi- and non-periodic
variables for easy reference. In addition, we present tables with the stars'
parameters. All light curves and full tables are available on request.

\acknowledgments
We are grateful to S. Koz{\l}owski and C. Kochanek for useful discussions and
a careful reading of the manuscript. We also thank the anonymous
referee for suggestions that greatly improved this paper.

DMS and KZS are supported in part by National Science Foundation grants
AST-0707982 and AST-0908816.
AZB acknowledges the support of the European Community under a Marie Curie
International Reintegration Grant.
GP and BP are supported by the Polish MNiSW grant N203 007 31/1328.
JLP acknowledges support from NASA through Hubble Fellowship grant
HF-51261.01-A awarded by the Space Telescope Science Institute, which is
operated by AURA, Inc. for NASA, under contract NAS 5-26555.

This research has made use of the Digitized Sky Survey images via the Goddard
{\it SkyView} utility (http://skyview.gsfc.nasa.gov/cgi-bin/query.pl), of
NASA's Astrophysics Data System, as well as the SIMBAD database, operated at
CDS, Strasbourg, France.


\clearpage
\bibliographystyle{apj}
\bibliography{ref}
\clearpage



\clearpage

\begin{figure*}
\begin{center}
\includegraphics[width=0.99\linewidth]{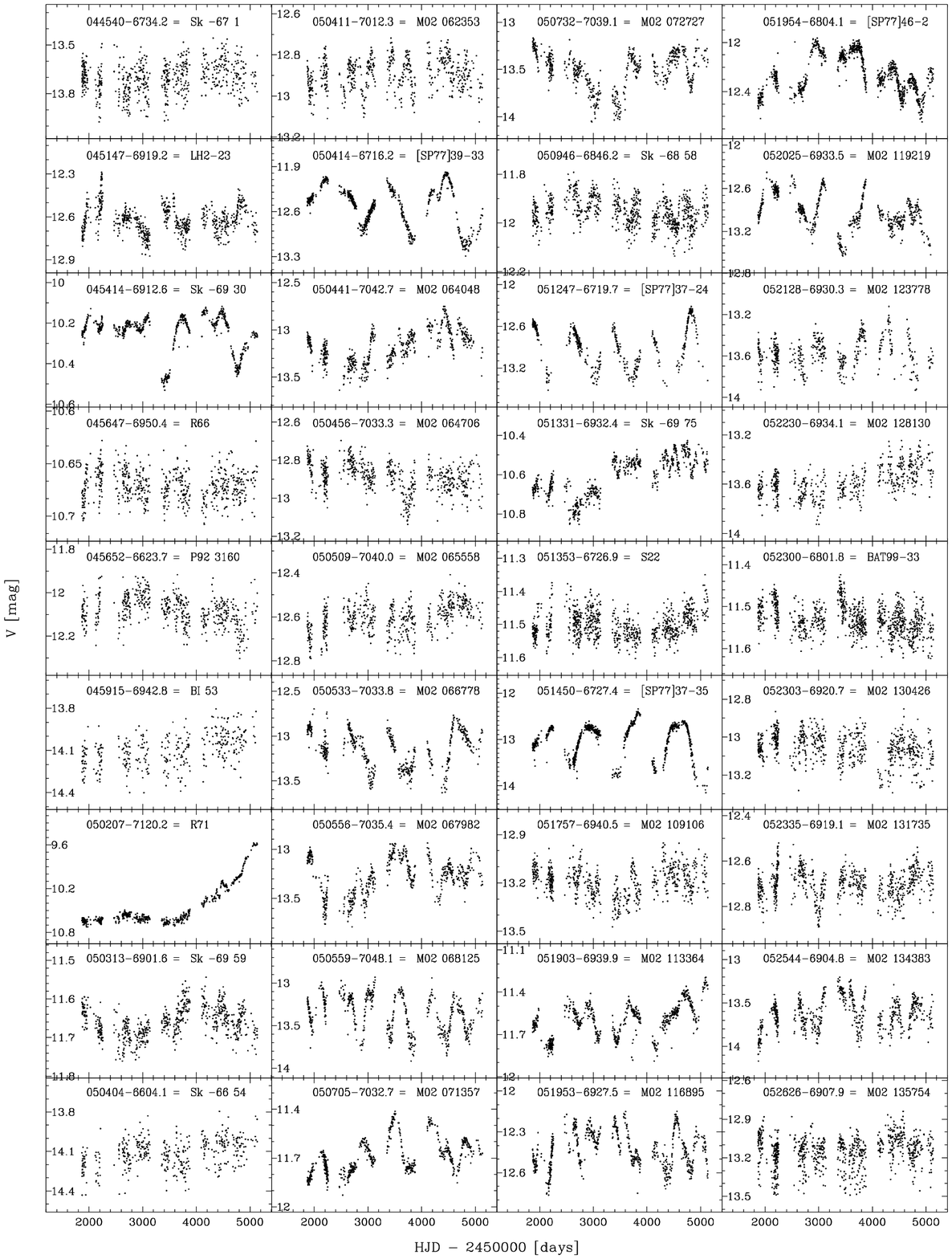} 
\end{center}
\caption{This and the following two pages present light curve stamps of all 104
semi- or non-periodic variable stars identified in the course of this study.}
\label{104_lc}
\end{figure*}

\begin{figure*}
\ContinuedFloat
\begin{center}
\includegraphics[width=0.99\linewidth]{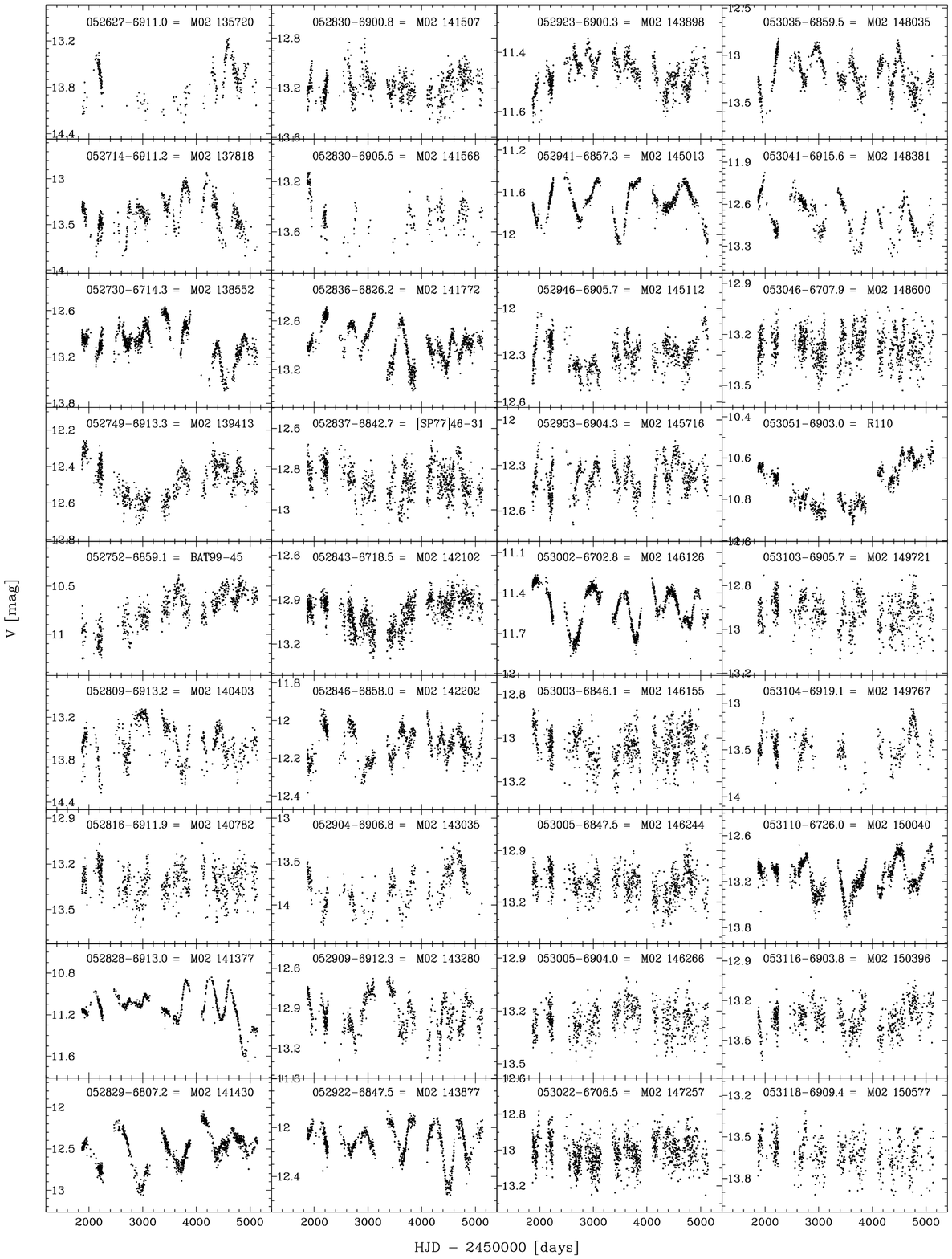}
\end{center}
\caption{(Continued)}
\label{104_lc}
\end{figure*}

\begin{figure*}
\ContinuedFloat
\begin{center}
\includegraphics[width=0.99\linewidth]{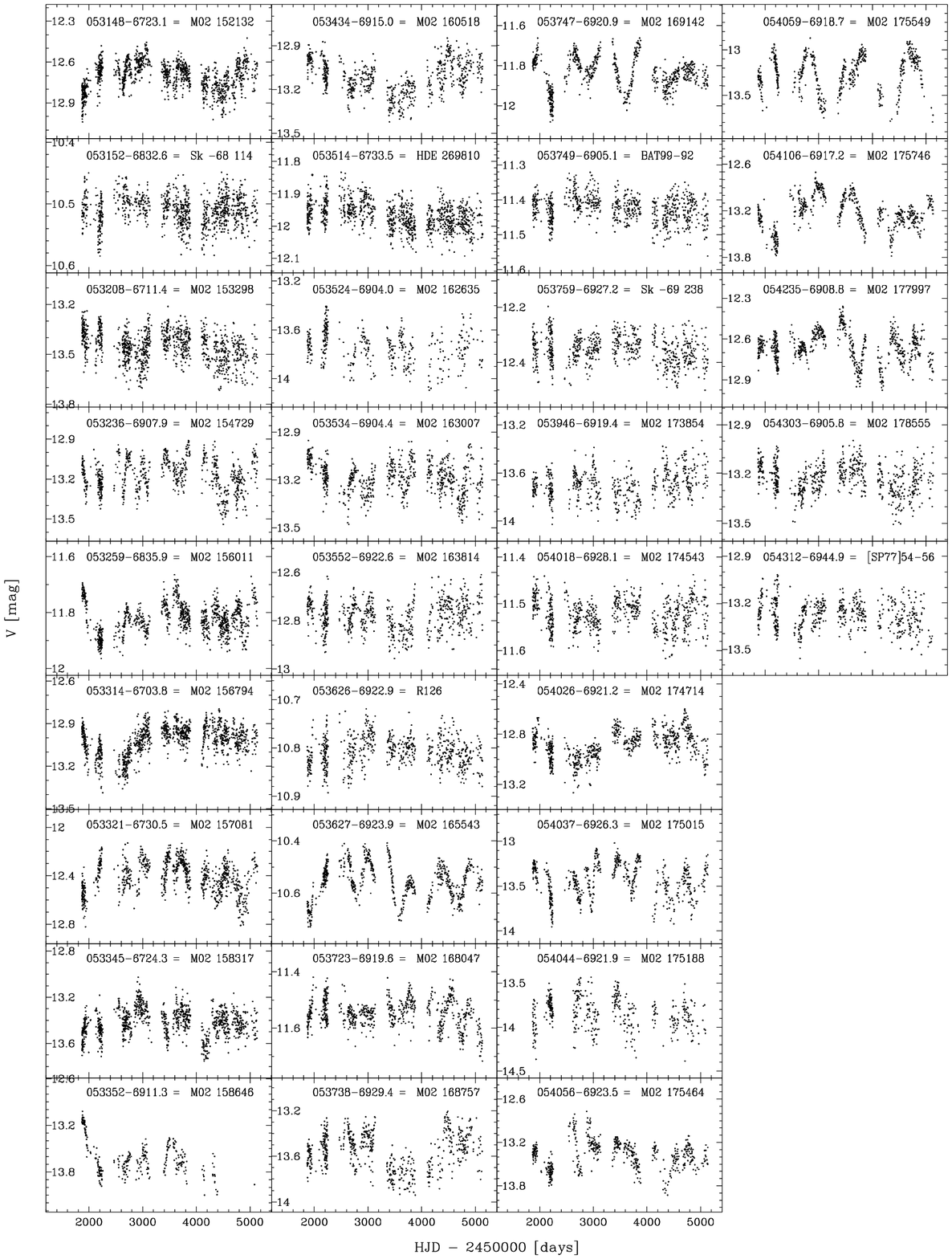}
\end{center}
\caption{(Continued)}
\label{104_lc}
\end{figure*}

\begin{figure*}
\begin{center}
\begin{tabular}{@{}c@{}c@{}}
\includegraphics[width=0.5\linewidth]{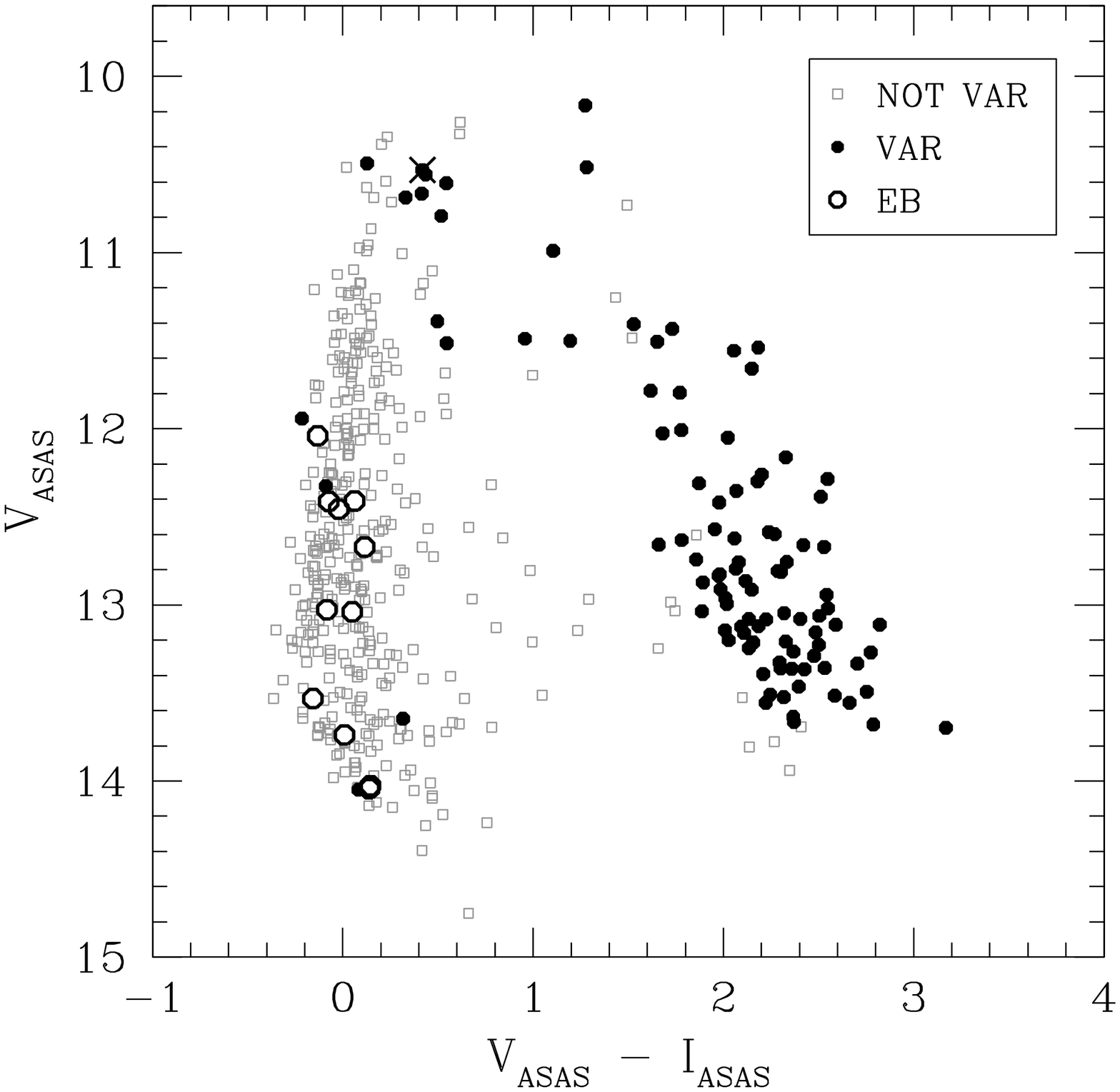} &
\includegraphics[width=0.5\linewidth]{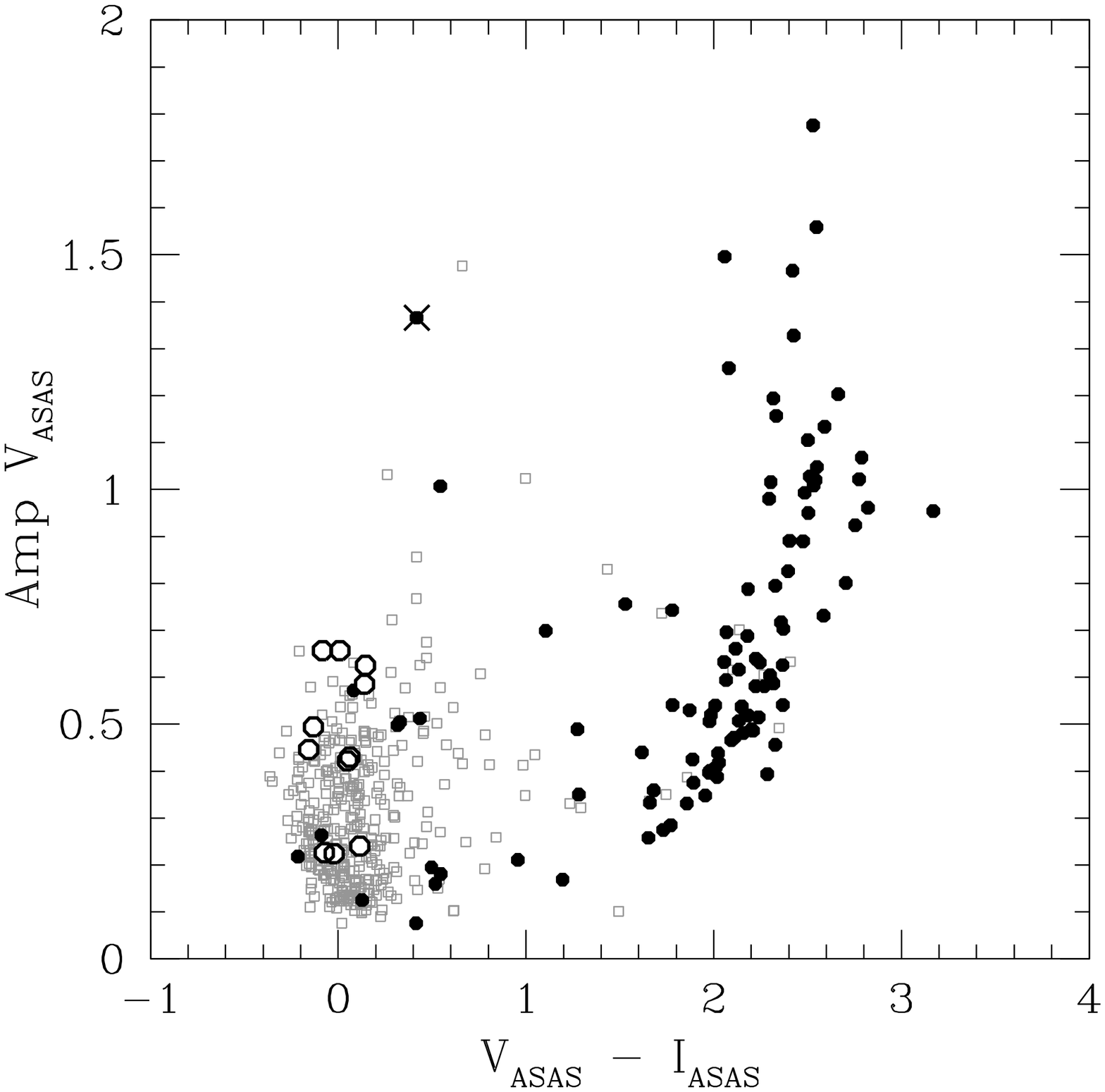} \\
\end{tabular}
\end{center}
\caption{Left panel: CMD for the 599 matches between ASAS and B09 catalogs.
$V$ and $I$ magnitudes are the mean values based on ASAS light
curves. Black dots represent the 117 variable stars and gray open squares the
non-variables. EBs are marked with big open circles.
Right panel: $V$ amplitude of 117 variables vs. $V-I$ color with the same
designations. For the non-variables, the light curve scatter is plotted.
R71 is marked with a cross (both panels).}
\label{vi}
\end{figure*}

\begin{figure*}
\begin{center}
\includegraphics[width=0.65\linewidth]{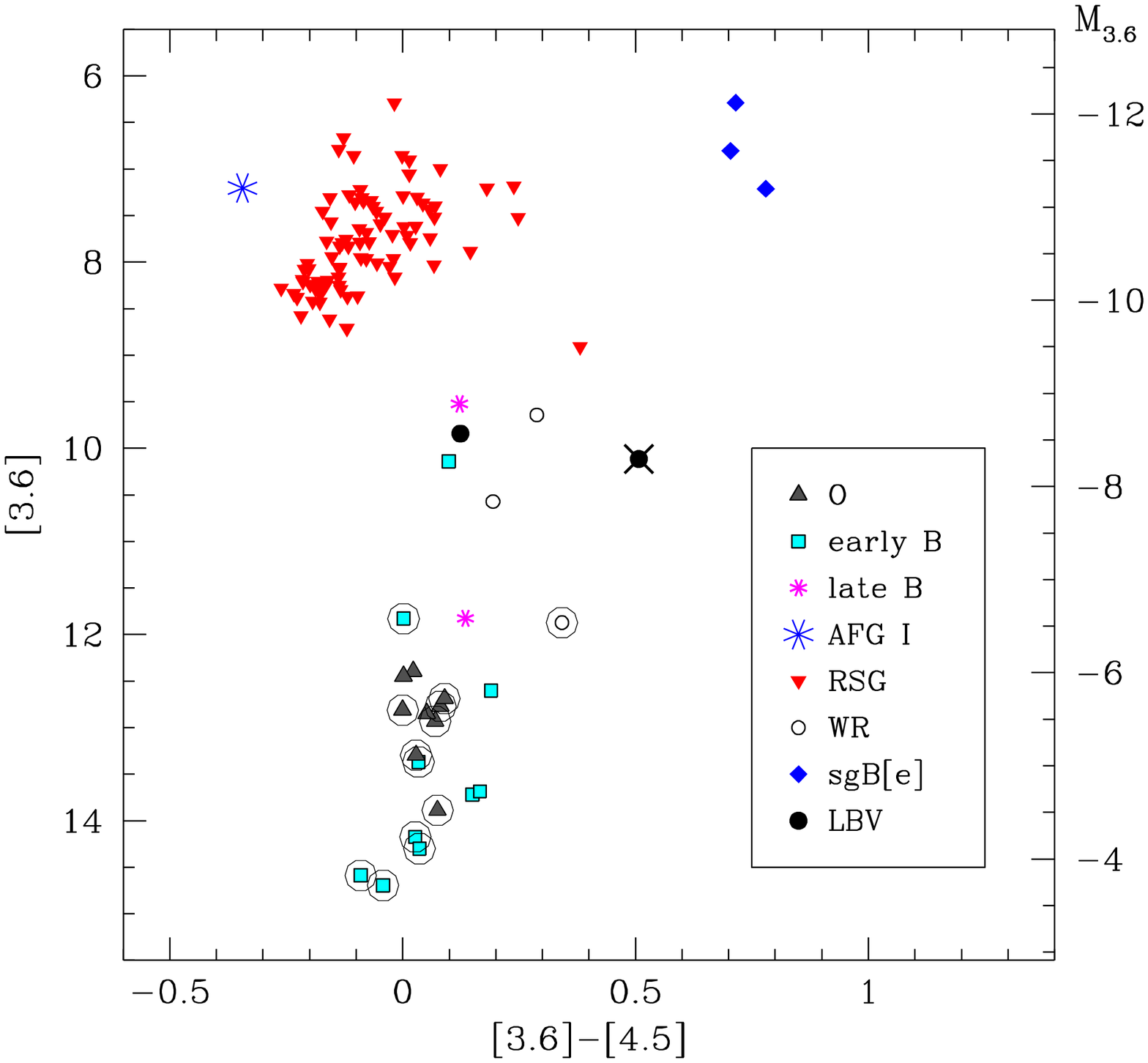}
\end{center}
\caption{Infrared CMD for 113 variable stars out of 599 matches between ASAS
and B09 catalogs, for which infrared colors were available. Values of
[3.6] and [4.5] were taken from the B09 catalog. The designations are the same
as in Figure 2 of B09. EBs are marked with big open circles and R71 is marked
with a cross.}
\label{36_45}
\end{figure*}
\clearpage



\begin{deluxetable}{lllccccl}
\tabletypesize{\scriptsize}
\tablecaption{Parameters of Known Binary Stars}
\tablewidth{0pt}
\tablehead{
\colhead{B09 ID\tablenotemark{a}} & \colhead{RA 2000} & \colhead{Dec 2000} &
\colhead{HJD$_0$ - 2450000} &
\colhead{Per} & \colhead{V$_{\rm ASAS}$} & \colhead{Amp$_V$} & \colhead{Sp Type} \\
\colhead{} & \colhead{[deg]} & \colhead{[deg]} & \colhead{[days]} & \colhead{[days]} & \colhead{[mag]} &
\colhead{[mag]} & \colhead{}
} 
\startdata
BAT99-6 (Sk -67 18)  & 73.8121643 &  -67.1899185 & 3060.318974 & 2.001248 & 12.04 & 0.49 & O3f* + O EB \\
BAT99-32 (Sk -71 21) & 80.5938721 &  -71.5994720 & 3425.880887 & 3.815192 & 12.67 & 0.24 & WN6(h) SB   \\
Sk -67 105           & 81.5257950 &  -67.1824417 & 3400.010490 & 3.301529 & 12.41 & 0.43 & O4 f + O6V        \\
Sk -67 117           & 81.8641281 &  -67.1984711 & 3412.698954 & 4.828888 & 13.03 & 0.66 & O9 III-II + O8 V: \\
MACHO81.8763.8       & 83.6720810 &  -69.5275040 & 3090.591666 & 1.404787 & 13.62 & 0.59 & O3 If* + O6:V     \\
\enddata
\tablenotetext{a}{Star designations following B09:
\citet[BAT99;][]{Breysacher99}, \citet[Sk;][]{Sanduleak70}}
\end{deluxetable}


\begin{deluxetable}{lllccccl}
\tabletypesize{\scriptsize}
\tablecaption{Parameters of Eight New Eclipsing Binaries}
\tablewidth{0pt}
\tablehead{
\colhead{B09 ID\tablenotemark{a}} & \colhead{RA 2000} & \colhead{Dec 2000} & 
\colhead{HJD$_0$ - 2450000} &
\colhead{Per} & \colhead{V$_{\rm ASAS}$} & \colhead{Amp$_V$} & \colhead{Sp Type} \\
\colhead{} & \colhead{[deg]} & \colhead{[deg]} & \colhead{[days]} & \colhead{[days]} & \colhead{[mag]} &
\colhead{[mag]} & \colhead{}}
\startdata
Sk -69 9   & 72.4642105 &  -69.2011948 & 3445.552851 & 1.736425 & 12.46 & 0.22 & O6.5 III  \\
Sk -70 18  & 74.0687943 &  -69.9936905 & 3445.140846 & 7.625148 & 13.04 & 0.42 & B0 n      \\
BI 98      & 78.1266632 &  -67.4539948 & 3245.926372 & 3.765188 & 14.03 & 0.62 & B1: V     \\
Sk -65 47  & 80.2277908 &  -65.4550781 & 3585.283137 & 3.701956 & 12.41 & 0.23 & O4 If     \\
Sk -71 29  & 81.2763748 &  -71.4644165 & 3462.286550 & 1.832248 & 14.04 & 0.58 & B extr    \\
Sk -67 102 & 81.5037918 &  -67.0528107 & 3450.143306 & 3.659535 & 13.53 & 0.45 & B2 III SB:\\
Sk -69 194 & 83.6503296 &  -69.7601395 & 3451.041843 & 12.22083 & 11.66 & 0.67 & B0 I + WN   \\
Sk -67 270 & 86.7945404 &  -67.9819183 & 3474.245086 & 3.067998 & 13.74 & 0.66 & B0.5 V    \\
\enddata
\tablenotetext{a}{Star designations following B09:
\citet[BI;][]{Brunet75}, \citet[Sk;][]{Sanduleak70}}
\end{deluxetable}



\begin{deluxetable}{lllcccclccl}
\tabletypesize{\scriptsize}
\tablecaption{Parameters of 117 Variable Stars Identified in this Study}
\tablewidth{0pt}
\tablehead{
\colhead{B09 ID\tablenotemark{a}} & \colhead{RA 2000} & \colhead{Dec 2000} & \colhead{V$_{\rm ASAS}$} & \colhead{Amp$_V$} &
\colhead{I$_{\rm ASAS}$} & \colhead{Amp$_I$} &  \colhead{Sp Type} & E(V-I)\tablenotemark{b} & \colhead{Known?\tablenotemark{c}} & 
\colhead{Neighbors\tablenotemark{d}} \\
\colhead{}  & \colhead{[deg]} & \colhead{[deg]} & \colhead{[mag]} & \colhead{[mag]} & \colhead{[mag]} &
\colhead{[mag]} & \colhead{} & \colhead{[mag]} & \colhead{} & \colhead{} }
\startdata
     P92 3160 &   74.2139130 &  -66.3970566 & 12.03 &  0.36 & 10.35 &  0.19 &  early K I &  ---  & --- & 1 bright, $\sim$25" \\
        BI 53 &   74.8157120 &  -69.7153549 & 14.04 &  0.61 &  ---  &  ---  &     B0 III &  ---  & --- & 1 bright, $\sim$20" \\
    Sk -69 59 &   75.8029175 &  -69.0269470 & 11.64 &  0.20 &  ---  &  ---  &O(f) +B1 I? & 0.000 & --- & 1 bright, $\sim$15" \\
   M02 062353 &   76.0413361 &  -70.2049942 & 12.84 &  0.40 & 10.86 &  0.15 &       M1 I & 0.120 & --- & --- \\
   M02 064706 &   76.2255402 &  -70.5552521 & 12.83 &  0.40 & 10.85 &  0.19 &     M0-1 I & 0.134 & --- & --- \\
   M02 065558 &   76.2917938 &  -70.6675568 & 12.57 &  0.35 & 10.62 &  0.16 &       M0 I & 0.000 & --- & --- \\
    Sk -68 58 &   77.4382935 &  -68.7694168 & 11.92 &  0.30 &  ---  &  ---  &     B3 Iab & 0.224 & --- & 1 bright, $\sim$25" \\
    Sk -69 75 &   78.3782501 &  -69.5399170 & 10.56 &  0.51 & 10.12 &  0.59 &       B8 I & 0.163 & V1577 & --- \\
   M02 119219 &   80.0987091 &  -69.5575867 & 12.76 &  1.26 & 10.68 &  0.91 &       M3 I & 0.126 & V2148 & --- \\
    Sk -65 47 &   80.2277908 &  -65.4550781 & 12.41 &  0.23 & 12.49 &  0.27 &      O4 If & --- & --- & --- \\
     BAT99-32 &   80.5938721 &  -71.5994720 & 12.67 &  0.24 & 12.56 &  0.37 &  WN6(h) SB & 0.141 & V2296 & --- \\
\enddata
\tablecomments{A full version of this table is available in the
electronic edition of the {\it Astronomical Journal}.}
\tablenotetext{a}{Star designations following B09:
\citet[BAT99;][]{Breysacher99}, \citet[Sk;][]{Sanduleak70},
\citet[W;][]{Westerlund61}, \citet[BI;][]{Brunet75},
\citet[LH;][]{Lucke72}, \citet[S;][]{Henize56}, \citet[P92;][]{Parker01},
\citet[M02;][]{Massey02}}
\tablenotetext{b}{Mean extinction values taken from \citet{Pejcha09}. Zero
values (0.000) correspond to negative values for the calculated extinction
A$_{I}$, which is the case for 30 out of 101 stars for which the extinction
value was available.}
\tablenotemark{c}{GCVS LMC designation.}
\tablenotemark{d}{Approximate brightness and distance of close neighbors,
based on DSS images. See Section 3.3 for details.}
\end{deluxetable}


\end{document}